\documentclass[aps,prb,preprint,superscriptaddress]{revtex4}

\usepackage{graphicx}
\usepackage{amsmath}

\begin{document}

\title{Prediction of Antiferromagnetism in Barium Chromium Phosphide 
Confirmed after Synthesis}

\author{R.A. Jishi}
\author{J.P. Rodriguez}
\affiliation{Department of Physics and Astronomy, 
California State University, Los Angeles, California 90032}
\author{T.J. Haugan}
\author{M.A. Susner}
\affiliation{Aerospace Systems Directorate, Air Force Research Laboratory, Wright-Patterson Air Force Base, OH 45433}


\begin{abstract}
We have carried out density-functional theory (DFT) calculations for 
the chromium pnictide BaCr$_2$P$_2$, which is structurally analogous to BaFe$_2$As$_2$, 
a parent compound for iron-pnictide superconductors. 
Evolutionary methods combined with DFT predict that the chromium analog has 
the same crystal structure as the latter. DFT also predicts
N\'eel antiferromagnetic order on the chromium sites. 
Comparison with
a simple electron-hopping model over a square lattice of chromium atoms 
suggests that it is due to residual nesting of the Fermi surfaces.
We have confirmed the DFT predictions directly after the
successful synthesis of polycrystalline samples of BaCr$_2$P$_2$. 
X-ray diffraction recovers the predicted crystal structure
to high accuracy, while magnetic susceptibility and specific-heat measurements 
are consistent with a transition to an
antiferromagnetically ordered state below $T_N \sim 60$ K.
\end{abstract}

\maketitle

\section{Introduction}
The discovery of iron-pnictide high-temperature superconductors represents one of 
the most important developments in condensed matter physics over the last ten years\cite{new_sc,paglione_greene_10}.  
Electric conduction in iron pnictides is due to the iron $3d$ electrons.
Elemental iron has six valence electrons in the $3d$ atomic shell, 
which is one more than half filled.
Elemental chromium, on the other hand,
has four valence electrons in the $3d$ atomic shell,
which is one less than half filled.
If Hund's rule is obeyed,
then elemental chromium is the particle-hole conjugate of elemental iron.
This observation has motivated a recent search 
for chromium analogs to iron-pnictide high-temperature superconductors.
In particular, the chromium-pnictide compound BaCr$_2$As$_2$ has been
synthesized\cite{singh_09,filsinger_17,richard_17,das_17}.
It has the same crystal structure as\cite{xtal_strctr} ThCr$_2$Si$_2$,
in common with BaFe$_2$As$_2$ 
and with other parent compounds to iron-pnictide superconductors.  
Also like parent compounds to iron-pnictide superconductors, 
BaCr$_2$As$_2$ is a bad metal that shows antiferromagnetic order on the chromium atoms.
Unlike the ``stripe'' antiferromagnetic order on the iron atoms 
shown by parent compounds to iron-pnictide superconductors, however,
BaCr$_2$As$_2$ shows N\'eel antiferromagnetic order
on the chromium atoms\cite{filsinger_17,das_17}.
All attempts to achieve superconductivity 
by injecting charge carriers into BaCr$_2$As$_2$
through chemical doping have failed so far\cite{filsinger_17}.

\begin{figure}
\includegraphics[scale=0.5]{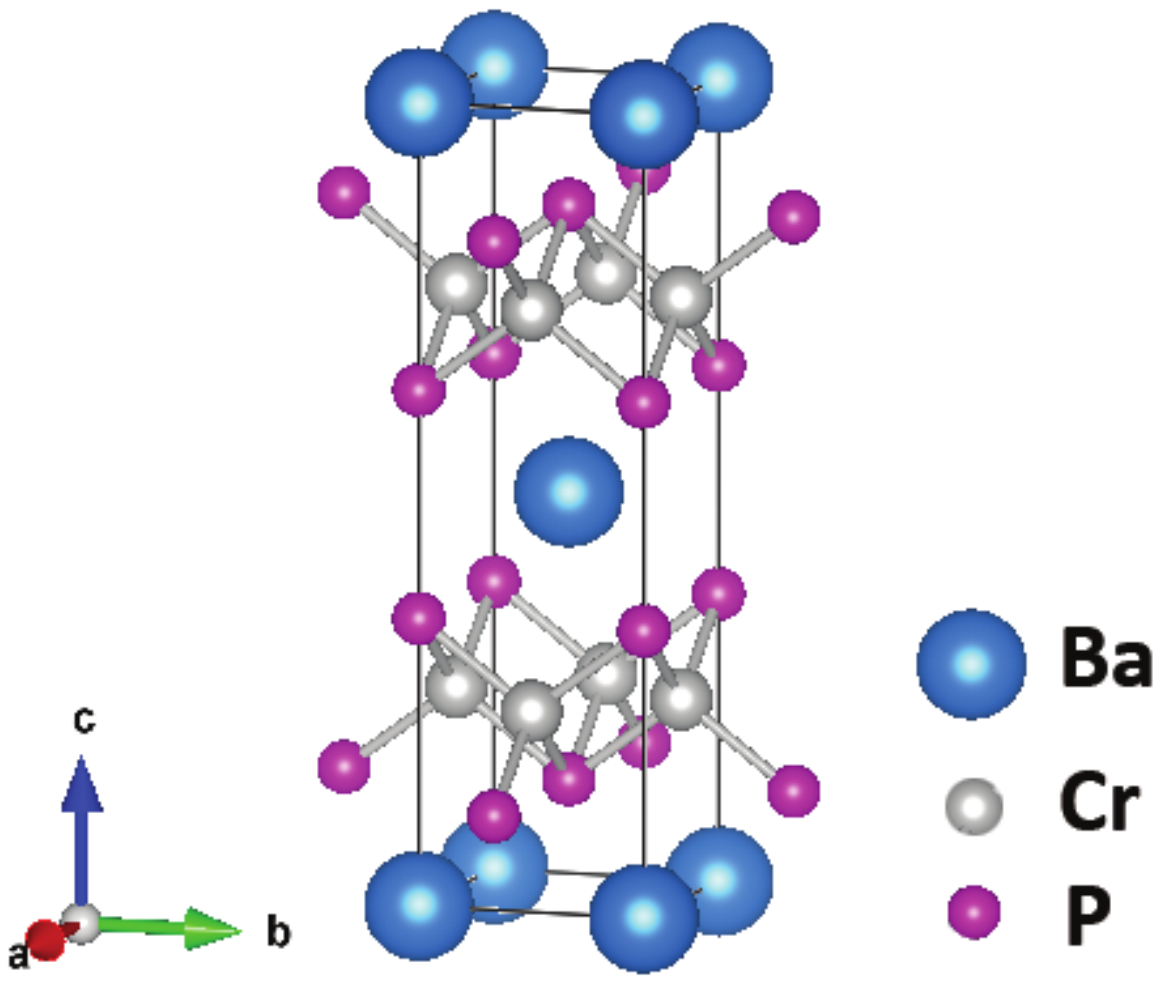}
\caption{Unit cell of BaCr$_2$P$_2$.}
\label{xtal_strctr}
\end{figure}

Synthesis of the chromium-pnictide sister compound BaCr$_2$P$_2$ 
has been recently reported\cite{bacr2p2}.
To our knowledge, however,
no report of the measured crystal structure of this new compound has been made in the literature.
This has motivated us to perform density-functional theory (DFT) calculations to
determine the electronic structure of the  new compound.
Biologically inspired 
optimization\cite{oganov_06,lyakhov_13,oganov_11}
of the crystal structure\cite{kresse_93,kresse_96,kresse_99}
results in the expected
ThCr$_2$Si$_2$-type crystal structure for the groundstate\cite{xtal_strctr},
but with N\'eel antiferromagnetic order on the chromium atoms.
(See Fig. \ref{xtal_strctr}.)
Analysis of a simple electron hopping model over a square lattice of chromium atoms 
that contain only the principal $3 d_{xz}$ and $3 d_{yz}$ orbitals
per atom suggests that this N\'eel order is due to residual nesting
of the Fermi surfaces
that is hidden by a Lifshitz transition of the latter.
We have also succeeded in synthesizing powder samples 
of the new compound BaCr$_2$P$_2$.  X-ray diffraction (XRD) on these powders
yields lattice constants that agree with our DFT
 predictions to within a percent.
Last, magnetic susceptibility and specific heat measurements are consistent with
a transition from a paramagnetic  to an antiferromagnetic state 
at temperatures below $T_N \cong 60$ K.

\section{Density-Functional-Theory Calculation}\label{DFT}
Below, we go into some detail on what evolutionary methods for structure predictions are,
followed by what it predicts for the electronic structure of BaCr$_2$P$_2$.

\subsection{Method}   
Theoretical prediction of the stable and metastable structures of
the compound BaCr$_2$P$_2$ was accomplished by using an
evolutionary scheme implemented in the code USPEX
(Universal Structure Prediction: Evolutionary Xtallography).
USPEX  was developed by 
Oganov, Glass, Lyakhov, and Zhu\cite{oganov_06,lyakhov_13,oganov_11};
it features local optimization, real space representation, 
and variation operators that mimic natural evolution. 

\begin{figure}
\includegraphics[scale=0.7]{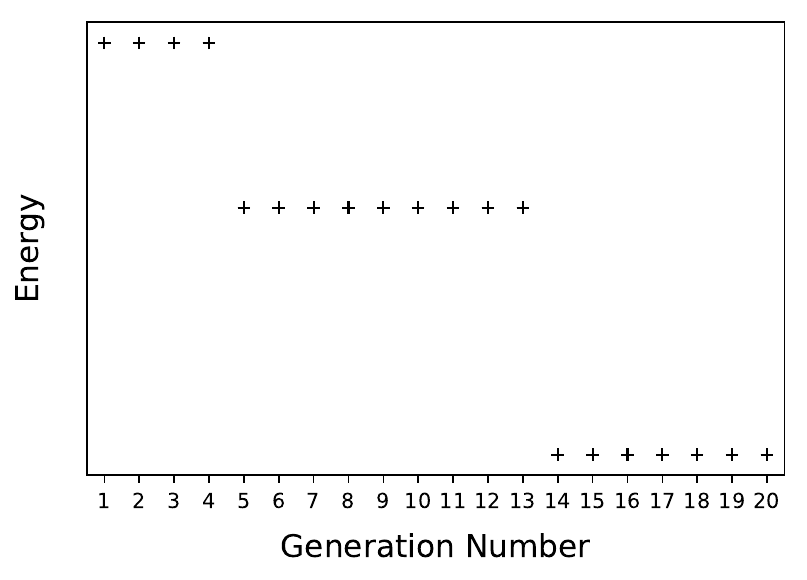}
\caption{The energy of the best structure predicted by USPEX as a function of the generation number.
In the last seven generations, the energy of the best strucure does not change.}
\label{generate}
\end{figure}

The method begins by generating a population of random crystal structures, 
each with a symmetry prescribed by a randomly chosen space group. 
For each structure, appropriate lattice vectors and atomic positions 
are generated in accordance with the selected space group. 
Density functional theory is then used to optimize the resulting structure 
and to calculate its free energy (known as the fitness function). 
Structure optimization is carried out 
using the VASP code \cite{kresse_93,kresse_96,kresse_99},
which uses a basis set of plane waves to expand the electronic wave function
and projector augmented wave (PAW) potentials.
Since the generated structures are usually far from equilibrium, 
the optimization procedure is carried out in four steps, 
beginning with a coarse optimization,
 which is followed by successively finer iterations. 
The set of the optimized structures of the 
initially generated population constitute the 
first generation. A new population of crystal structures is 
then produced, some members of which being randomly generated, 
while others are obtained as offspring  of the best structures 
(those with lowest energy) of the previous generation. 
The offspring are derived from parent structures 
by applying variation operators such as heredity, mutation, or permutation. 
In each generation, 30\% of the structures are generated randomly,
40\% by heredity (one structure produced from two parent structures),
10\% by soft mutation (atom movements along the softest mode),
10\% by lattice mutation (variation of lattice constants),
and 10\% by permutation (exchanging atoms of different types).
The optimized structures in the new population form the second generation, 
and the best among them serve as precursors for a new generation. 
The process continues until convergence to the most stable structures is attained.
In applying the evolutionary method, an initial population of 200 structures
constitutes the first generation; subsequent generations consist of 40 structures each.
Convergence to the best structures was achieved after evaluating 960 structures.
The structural stability of the best predicted structure is checked by 
calculating its spectrum of phonon frequencies, using density functional perturbation theory
as implemented in Quantum Espresso\cite{QE}.  A grid of $4\times 4\times 4$
$q$-vectors is used to reconstruct the dynamical matrix in real space
and subsequently to calculate the phonon frequencies.

As stated above, structure optimization is carried out in four steps
that gradually increase in accuracy.
In the first, second, third, and fourth steps, the tolerance for the convergence
of the self-consistent electronic loop is given by 
$0.05$ eV, $0.005$ eV, $0.001$ eV, and $0.0001$ eV, respectively.
The corresponding values for the ionic relaxation loop are
$0.5$ eV, $0.05$ eV, $0.01$ eV, and $0.001$ eV.
The Brillouin zone is sampled with a grid spacing given by $2\pi s$ \AA$^{-1}$,
where the grid becomes finer in successive optimization steps, with the parameter $s$
respectively given by $0.12$, $0.10$, $0.08$, and $0.05$ in the first, second, third, and fourth step.
The cutoff energies for the plane waves basis set 
are successively given by $300$ eV, $350$ eV, $420$ eV, and $500$ eV for the four optimization steps.
The energy of each optimized structure is finally determined via a static calculation,
where the cutoff energy is set to $600$ eV.

For the most stable structure, energy bands and densities of states 
were calculated using the all-electron, full-potential, linearized, 
and augmented plane wave method\cite{blaha_01}.
Here, space is divided into two regions; 
one region consists of the interior of non-overlapping spheres 
(known as muffin-tin spheres) centered at the atomic sites,
while the rest of space (the interstitial) 
forms the other region. The radii of the muffin-tin spheres centered on Ba, Cr, and P were
chosen to be $2.50$ a.u., $2.44$ a.u., and $2.00$ a.u., respectively.  The basis set used to expand 
the electronic wave function consists of plane waves in the interstitial, 
with a maximum wave vector of magnitude $K_{max}$, 
and where each plane wave is augmented by an atomic-like function 
in each muffin-tin sphere. 
The wavenumber $K_{max}$ was chosen so that $K_{max} R_{mt} = 8$, where $R_{mt}$ is 
the radius of the smallest muffin-tin sphere in the unit cell. 
Charge density was Fourier-expanded up to a maximum wave vector of $14 a_0^{-1}$,
where $a_0$ is the Bohr radius. 
For total energy calculations, a $15\times 15\times 15$ grid of $k$-points was used 
to integrate functions over the Brillouin zone, 
and convergence of the self-consistent field calculations 
was achieved with an energy tolerance of $0.0001$ Ry and 
a charge tolerance of $0.001 e$.

\begin{table}
\begin{tabular}{|c|c|c|c|}
\hline
space group (no.) & lattice constants (\rm\AA) & angles (degrees) & energy/atom (eV) \\
\hline
I4/mmm (139) & $a = b = 3.843$, $c = 13.300$ & $\alpha = \beta = \gamma = 90$ & $0.000$ \\
Cm (8) & $a = 12.667$, $b = 12.367$, $c = 4.138$ & $\alpha = \beta = 90$, $\gamma = 160$.9 & 0.068 \\
P4/mmm (123) & $a = b = 3.847$, $c = 6.841$ & $\alpha = \beta = \gamma = 90$ & $0.116$ \\
\hline
\end{tabular}
\caption{
The most stable crystal structures of BaCr$_2$P$_2$ 
predicted by evolutionary methods
at atmospheric pressure and at zero temperature. 
The space groups, lattice constants, angles between lattice vectors,
and relative energies per atom are listed. 
For the most stable structure, the fractional coordinate of the phosphorous atom
along the $c$-direction is $z = 0.35319$ \AA.
To facilitate comparison, the energy per atom of 
the most stable structure is set equal to zero.}
\label{table1}
\end{table}

\subsection{Predictions}
\begin{figure}
\includegraphics[scale=0.5,angle=90]{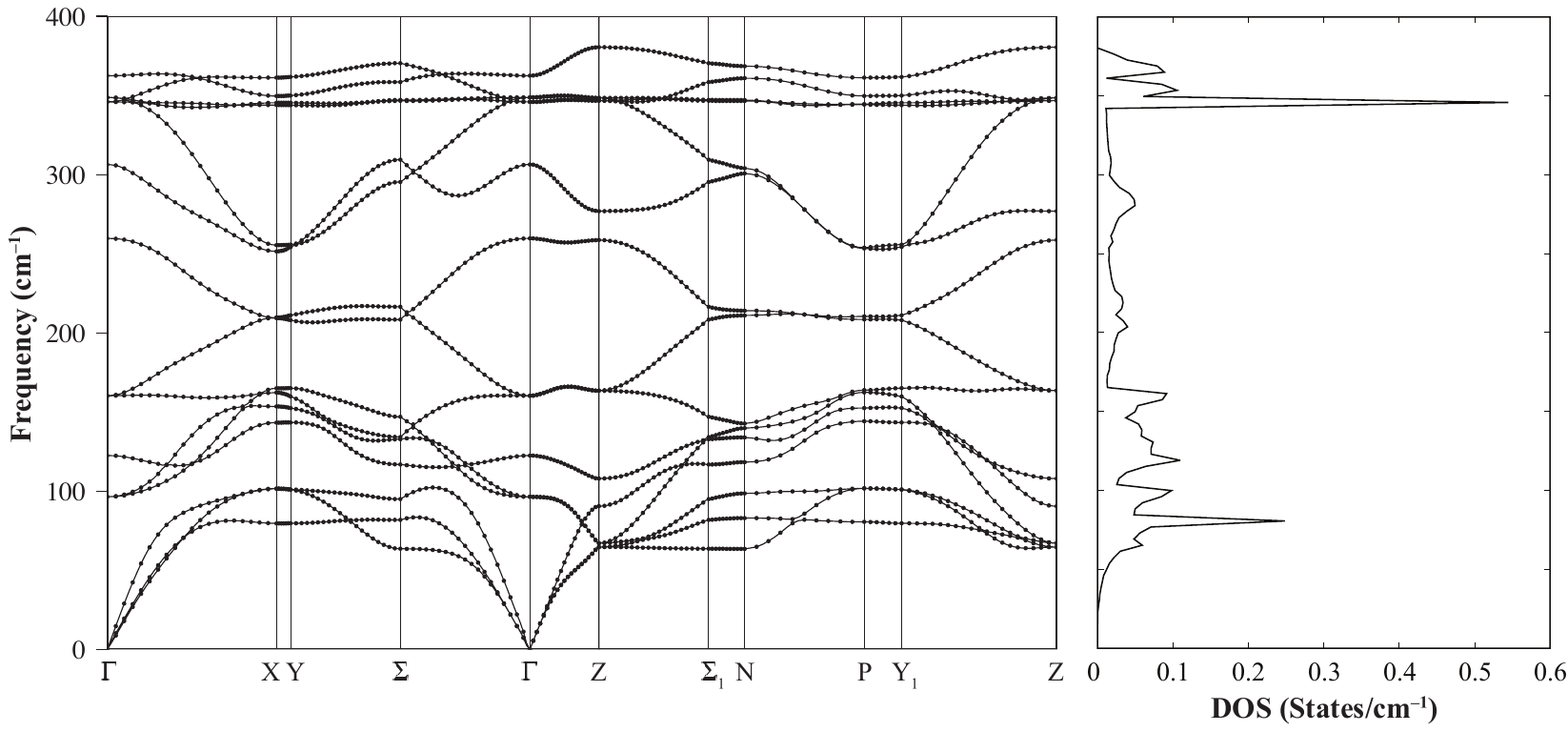}
\caption{Phonon dispersion curves and density of states predicted by density functional perturbation theory.}
\label{phonons}
\end{figure}

Figure \ref{generate} is a plot of the energy of the best structure,
predicted by the evolutionary methods, versus the generation number.
In the last seven generations, the energy of the best strucuture is
about $1$ meV lower than in previous generations.
Table \ref{table1} lists
the most stable states for BaCr$_2$P$_2$ on the basis of evolutionary methods combined with DFT.
We should note that in applying the evolutionary methods, we are looking
for the best crystal structure of the stoichiometric compound BaCr$_2$P$_2$.
The groundstate has the ThCr$_2$Si$_2$ crystal structure\cite{xtal_strctr}
shown in Fig. \ref{xtal_strctr},
which is  common
to the iron and chromium arsenides BaFe$_2$As$_2$ and BaCr$_2$As$_2$.
As shown later, the predicted most stable structure is in excellent agreement with
that determined experimentally.  To determine the magnetic order in the groundstate,
we calculated the energies of structures with different magnetic configuration,
using the experimental lattice constants.  The results are shown in Table \ref{table2}.
AFM stands for antiferromagnetic, FM for ferromagnetic, and NM for nonmagnetic.
In the A-AFM state, the magnetic order is ferromagnetic within the $a$-$b$ plane
and antiferromagnetic between the $a$-$b$ planes.  In the C-AFM state,
the order is ferromagnetic between parallel $a$-$b$ planes,
but antiferromagnetic within the planes, while in the G-AFM state,
the order is antiferromagnetic both within the $a$-$b$ planes and between the adjacent planes.
We conclude that in the groundstate,
the magnetic order is three-dimensional (3D) N\'eel on the chromium atoms.
The calculated magnetic moment is approximately $2.93$ Bohr magnetons per chromium atom.
The magnetic groundstate according to our DFT calculations is then a
spin-$1$ N\'eel antiferromagnet on the chromium atoms.
Figure \ref{phonons} presents the calculated phonon dispersion curves
along high-symmetry directions in the Brillouin zone, along with the phonon density of states.
The absence of imaginary frequencies is indicative of the stability of the crystal structure.

\begin{table}
\begin{tabular}{|c|c|}
\hline
magnetic order & energy per formula unit (eV) \\
\hline
G-AFM &  0.000  \\
C-AFM &  0.016  \\
A-AFM &  0.475  \\
FM    &  0.447  \\
NM    &  1.620  \\
\hline
\end{tabular}
\caption{Relative energies per formula unit for various magnetic orders in BaCr$_2$P$_2$. Here, NM
means  nonmagnetic, FM ferromagnetic, and AFM antiferromagnetic.  In A-AFM, the order is FM within
the $a$-$b$ planes, but AFM between adjacent planes.  In C-AFM, the order is AFM within
the $a$-$b$ planes, but FM between adjacent planes.  In G-AFM, the order is AFM both within
the $a$-$b$ planes and between adjacent planes.}
\label{table2}
\end{table}

\begin{figure}
\includegraphics[height=10cm,width=0.5\textwidth,angle=0]{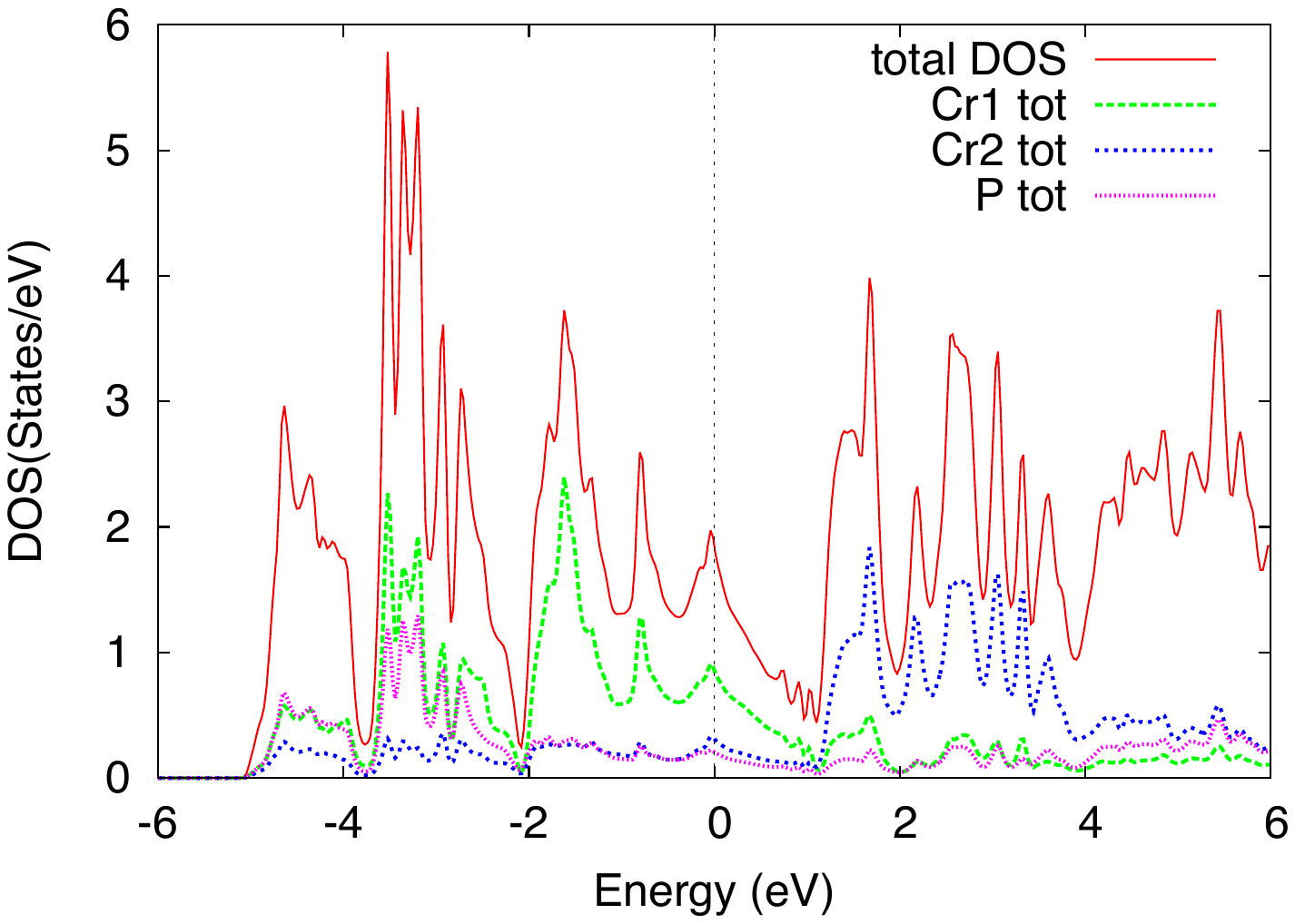}
\caption{Density of states of BaCr$_2$P$_2$ predicted by DFT.}
\label{DoS}
\end{figure}

Figure \ref{DoS} shows the energy bands and the density of states predicted by DFT for BaCr$_2$P$_2$.
The Fermi level lies at zero units of energy.  
We shall now follow the arguments made by Singh et al. in ref. \cite{singh_09}
for the case of the sister compound BaCr$_2$As$_2$.
Notice that the contribution to the density of states that remains after subtracting off 
the contributions from both the chromium majority-band (Cr1) and the chromium minority-band (Cr2) 
is sizable at the Fermi level.
It is approximately $1/3$ of the total.
The remaining contribution to the density of state originates
from the  phosphorus $3p$ orbitals, which are extended.
The latter property accounts for why the contribution due to the phosphorus atoms
shown in Fig. \ref{DoS} is much smaller,
where the extended $3p$ states are projected onto the
small  volume of the phosphorus atoms inside a unit cell.
We conclude that 
like in the case of the sister compound\cite{singh_09} BaCr$_2$As$_2$,
a sizable amount of hybridization
between the chromium $3d$ states and the phosphorus $3p$ states
exists at the Fermi level.
This is unlike what occurs in parent compounds 
to iron-pnictide superconductors such as BaFe$_2$As$_2$, 
in which case the  orbital character 
near the Fermi level is primarily due to the iron $3d$ levels
because the pnictogen $3p$ levels are localized\cite{singh_du_08,singh_09}.

\begin{figure}
\includegraphics[height=10cm,width=0.5\textwidth]{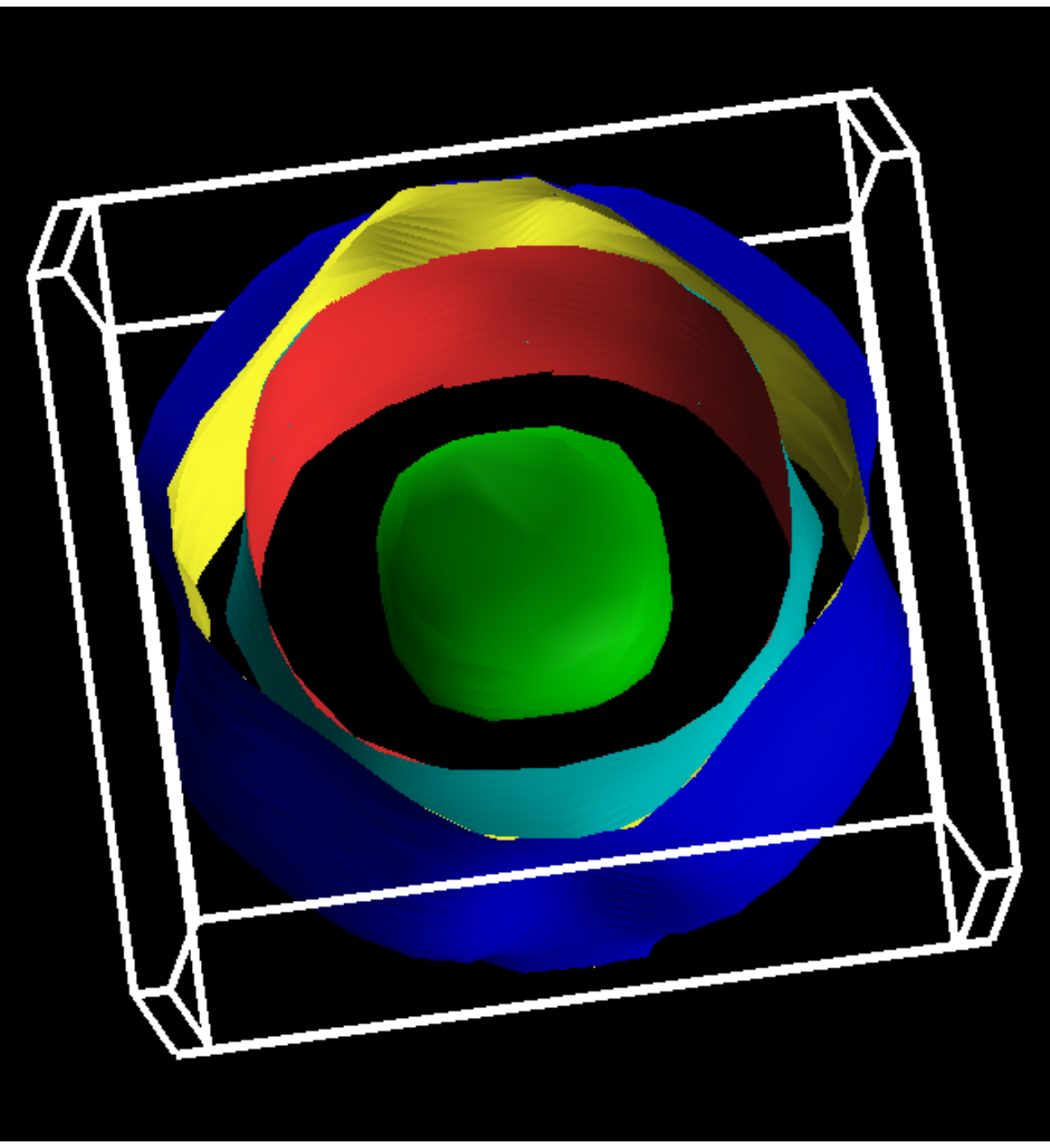}
\caption{Fermi surfaces of BaCr$_2$P$_2$ predicted by DFT.}
\label{FS}
\end{figure}

\begin{figure}
\includegraphics[scale=0.5]{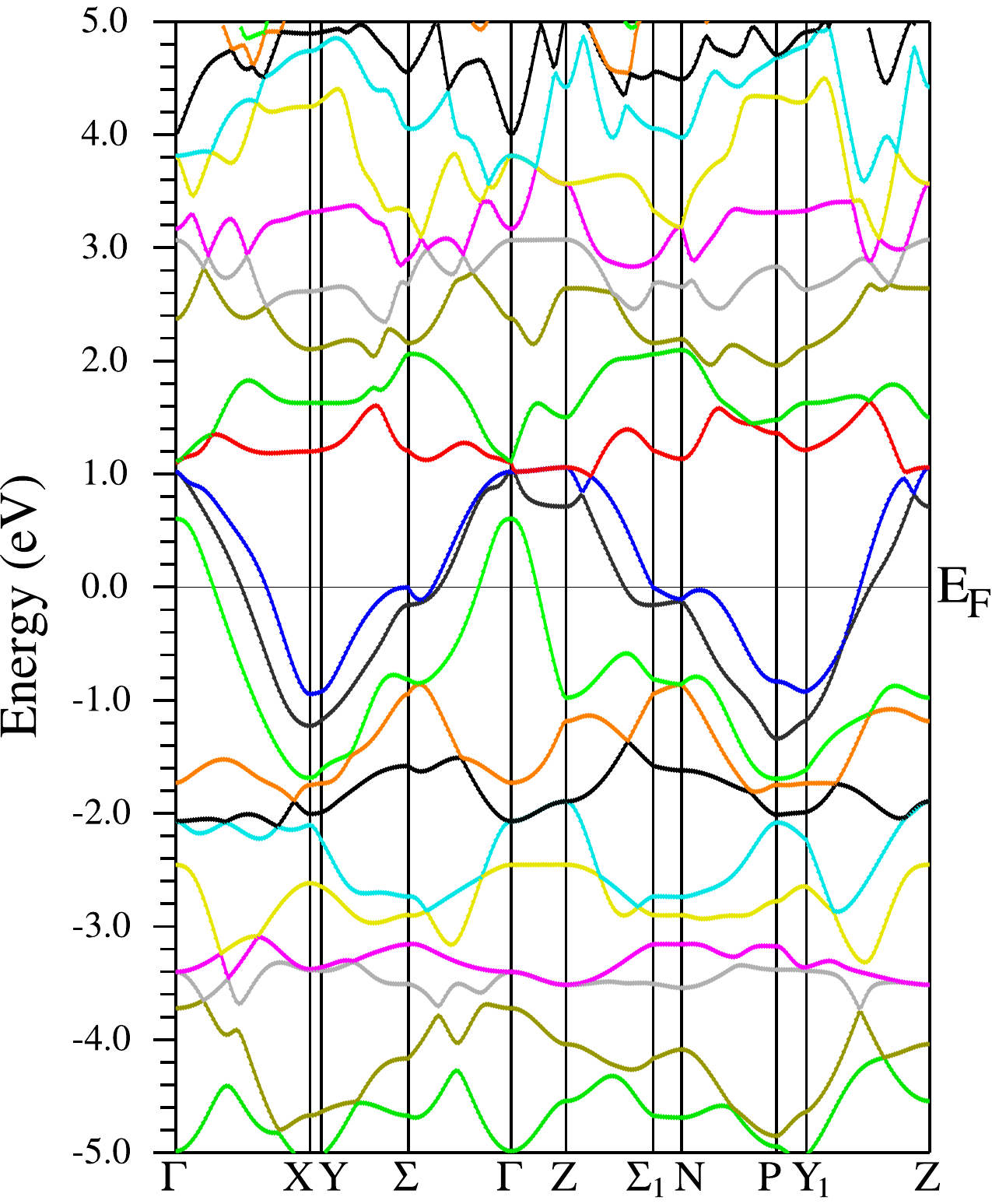}
\caption{Structure of majority bands in N\'eel-ordered BaCr$_2$P$_2$,
as predicted by DFT.}
\label{bnd_strctr}
\end{figure}

Last, Fig. \ref{FS} shows the Fermi surfaces predicted for the N\'eel groundstate
by DFT.  They are characterized 
by a 3D Fermi surface pocket centered at the $\Gamma$-point
and two tubular Fermi surface sheets along the  $c$ axis.
Inspection of the band structure
shown in Fig. \ref{bnd_strctr} 
reveals that all three Fermi surface sheets are hole-type.
The outer tube shows considerable dispersion along the $c$ axis.
These Fermi surfaces resemble
those predicted earlier for chromium arsenide by DFT\cite{singh_09}.

\section{Experimental Procedures}
Below, we report on the successful synthesis of BaCr$_2$P$_2$, 
as well as on X-ray diffraction on the resulting powder samples. 
Comparison with the previous predictions by DFT 
of a ThCr$_2$Si$_2$-type crystal structure 
and of a N\'eel antiferromagnetic groundstate over the chromium atoms is also made. 
\subsection{Materials Synthesis}
The compound BaCr$_2$P$_2$ was synthesized using conventional techniques. 
First, stoichiometric quantities of Ba (Alfa Aesar, Dendritic, 99.9\,\%),
 Cr (Alfa Aesar, powder -10+20 mesh, 99.996\,\%),
 and P (Alfa Aesar, lump, red 99.999\,\%) 
were combined in a stoichiometric quantity inside of a 2 mL volume Al$_2$O$_3$ crucible
and subsequently sealed under vacuum into a quartz ampoule. 
Prior to use, the Cr powder was reduced under flowing forming gas (95\,\% Ar, 5\,\% H$_2$)
for 12 hours. The ampoule was then placed inside of a box furnace and heated to $500 ^{\circ}{\rm C}$ over
a period of 24 hours and held at this temperature for the same period of time,
followed by a ramp-up to $1000 ^{\circ}{\rm C}$ over 17 hours and a dwell at this last temperature 
for 24 hours. After this last heat-treatment step, the ampoule was allowed to 
furnace-cool to room temperature. The mixture was extracted from the crucible, 
ground and pressed into a pellet in an Ar-filled glovebox, placed in 
a 2 mL Al$_2$O$_3$ crucible, and sealed into a quartz ampoule under 1/3 atm Ar. 
In this second anneal, the ampoule was taken to $1000 ^{\circ}{\rm C}$ over a period of 10 hours 
and held at this temperature for 24 hours, followed by a furnace cool.

The uniformly grey pellet was removed from the ampoule in an Ar-filled glove box
and broken into three sections for different 
materials characterization investigations.  
We found that the material garnered a greyish-white coating 
when exposed to air for periods longer than a couple of hours, so we 
endeavored to minimize exposure as much as possible. 

We characterized the structure of the material through 
X-ray diffraction using a Bruker D8 DaVinci system. 
In these experiments, we used Co radiation (Co $k_{\alpha} = 1.789190\, {\rm \AA})$. 
We also performed high-temperature diffraction under vacuum using a 
Perkin Elmer DHS 1100 heater with a graphite dome. 
FullProf software\cite{FullProf} was used to perform Rietveld refinement
on the resultant diffraction patterns. 
The material was relatively phase-pure ($\sim$ 97\,\%) and 
contained only a slight amount of the CrP impurity. 

Magnetic and thermal measurements were performed using 
a Quantum design Model 6000 Physical Property Measurement System (PPMS). 
The former set of measurements used the VSM option, where we used
the profiles of the magnetization versus both the temperature $T$ and
the magnetic field $H$ 
to elucidate the fundamental properties of the polycrystalline sample. 
In these measurements, we performed a zero-field cool operation before 
all magnetization versus temperature measurements. 
The thermal measurements were performed using the heat capacity option. 
In these measurements, we used the standard pulsed calorimetry option with 2\,\% 
temperature rise for the full temperature range and 30\,\% rise near
the phase transition to better characterize the subtle peak we found to be present.

\begin{figure}
\includegraphics[scale=0.5]{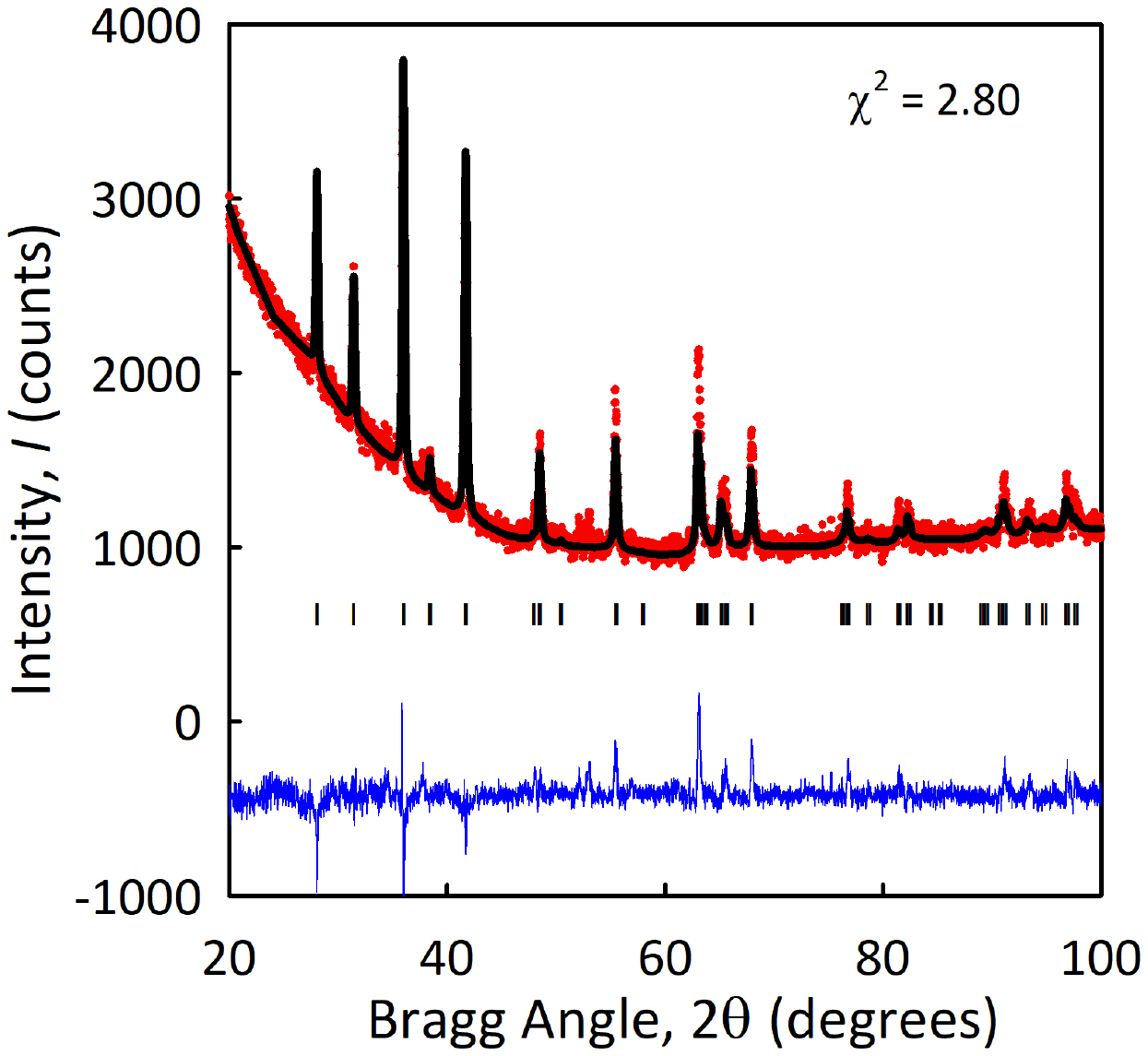}
\caption{Rietveld fit to the XRD pattern 
(Co $k_{\alpha}  = 1.789190 {\rm \AA}$) taken for BaCr$_2$P$_2$. 
The red dots represent the collected pattern, the black line is 
the fit based on the refinement, and the blue line is the difference plot. 
The 
hatch marks represent expected peak positions for the refined structure.}
\label{xrd}
\end{figure}

\subsection{Crystal Structure}
We applied full-pattern fitting to our experimental X-ray diffraction data using
the FullProf data analysis software package\cite{FullProf}. 
The known structure for BaFe$_2$P$_2$ was used as a starting point for 
our investigation\cite{Mewis_80} by simply substituting the Cr for the Fe atoms. 
The resulting fit was of good quality, and it is presented in Fig. \ref{xrd}.
With respect to the atomic coordinates within the unit cell, 
the P position was the only degree of freedom available for fitting. 
We present the results from the Rietveld analysis in Table \ref{table3}.

\begin{table}
\begin{tabular}{|c|c|c|c|}
\hline
site   & $x_{atom}$  & $y_{atom}$  & $z_{atom}$  \\
\hline
barium atom & $0$ & $0$ & $0$ \\
chromium atom & $1/2$ & $0$ & $1/4$ \\
phosphorus atom  & $0$ & $0$ & $0.35782(88)$ \\
\hline 
\hline 
$R_p$  &  $R_{wp}$  &    $R_{exp}$    & $\chi^2$ \\
\hline
$3.10$   &   $4.46$   &    $2.67$    &  $2.80$  \\
\hline
\end{tabular}
\caption{
Atomic coordinates and reliability factors from
Refined Crystal Structure of BaCr$_2$P$_2$ (I4/mmm, No. 139) at $300$ K,
where $a=b=3.8472(2)\, {\rm \AA}$ and $c=13.220(12)\, {\rm\AA}$.}
\label{table3}
\end{table}

\begin{figure}
\includegraphics[scale=0.5]{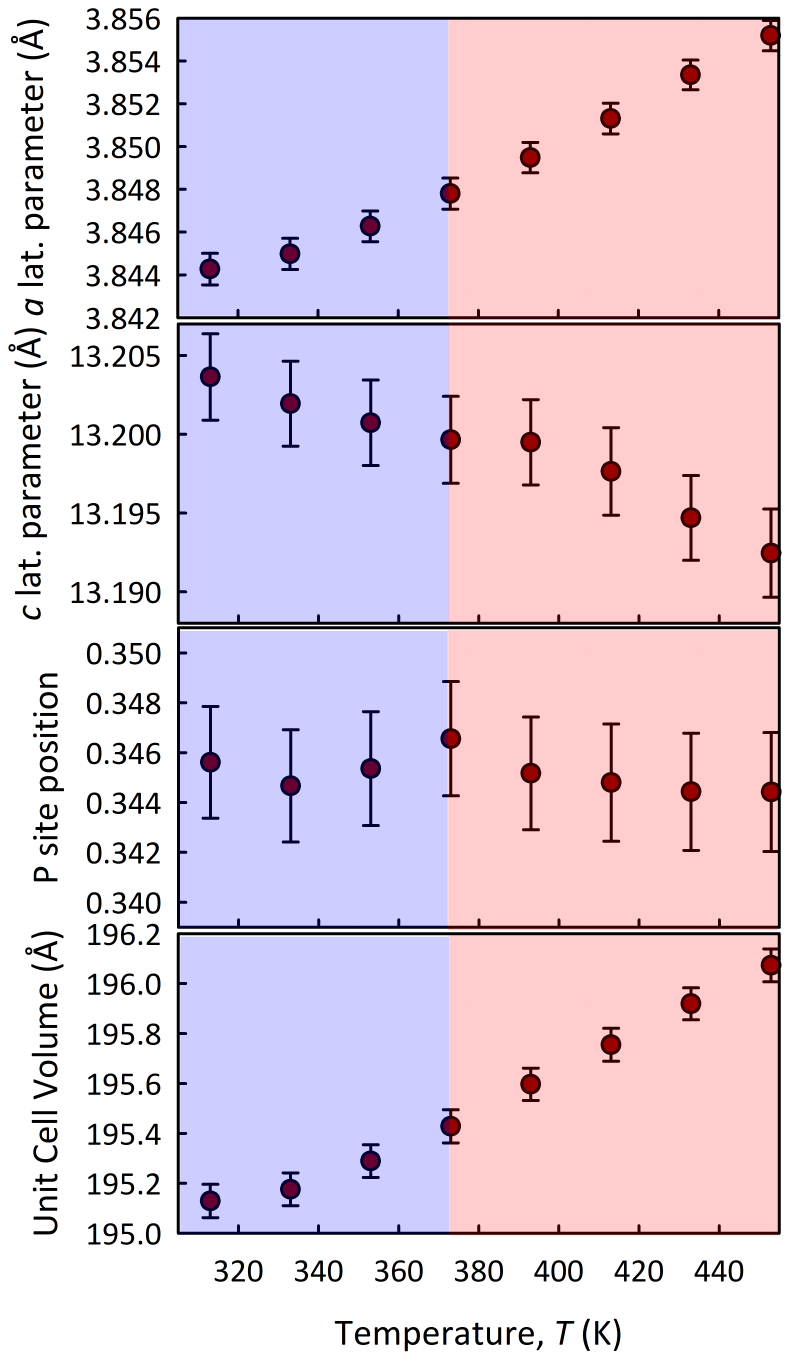}
\caption{Evolution of structural parameters with temperature.}
\label{xrd_temp}
\end{figure}

Like its sister compound\cite{singh_09,filsinger_17,das_17} BaCr$_2$As$_2$, 
the crystal structure of BaCr$_2$P$_2$ 
is of the ThCr$_2$Si$_2$ type shown in Fig. \ref{xtal_strctr}.
Further, the crystal structure determined by X-ray diffraction is in excellent agreement
 with that predicted by DFT, 
with an error in the lattice constants of less than one percent.  
It is important to mention that the correct crystal structure predicted by DFT 
(Table \ref{table1}) preceded our synthesis of BaCr$_2$P$_2$ and the subsequent XRD analysis
(Table \ref{table3}).

We performed high-temperature XRD to determine if
any of the anomalies in the magnetization measurements and in the specific heat measurements
to be reported below
could be ascribed to changes in crystal structure. 
Figure \ref{xrd_temp} shows the Rietveld refinements of this experiment.
Generally, as temperature is increased, 
the total thermal expansion is dominated by an increase in the $a$-lattice parameter.
At the same time, however, there is a small anomaly in the evolution of the $c$-lattice parameter
with temperature that coincides with a small anomaly in
the specific heat data reported below. 
This shift also coincides  with a subtle peak in the evolution of the $z$ position of the P atom.

\begin{figure}
\includegraphics[scale=0.6]{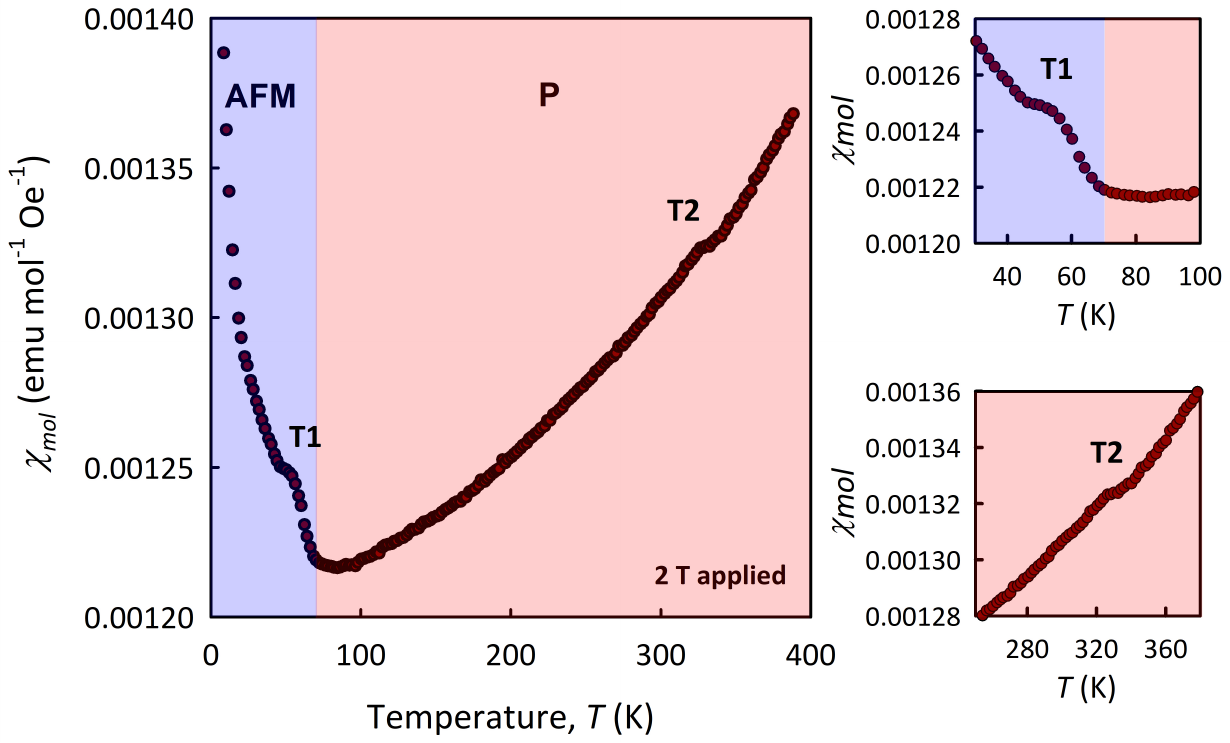}
\caption{Magnetic susceptibility versus Temperature for BaCr$_2$P$_2$.}
\label{chi_vs_T}
\end{figure}

\subsection{Magnetization and Specific Heat}
\begin{figure}
\includegraphics[scale=0.5]{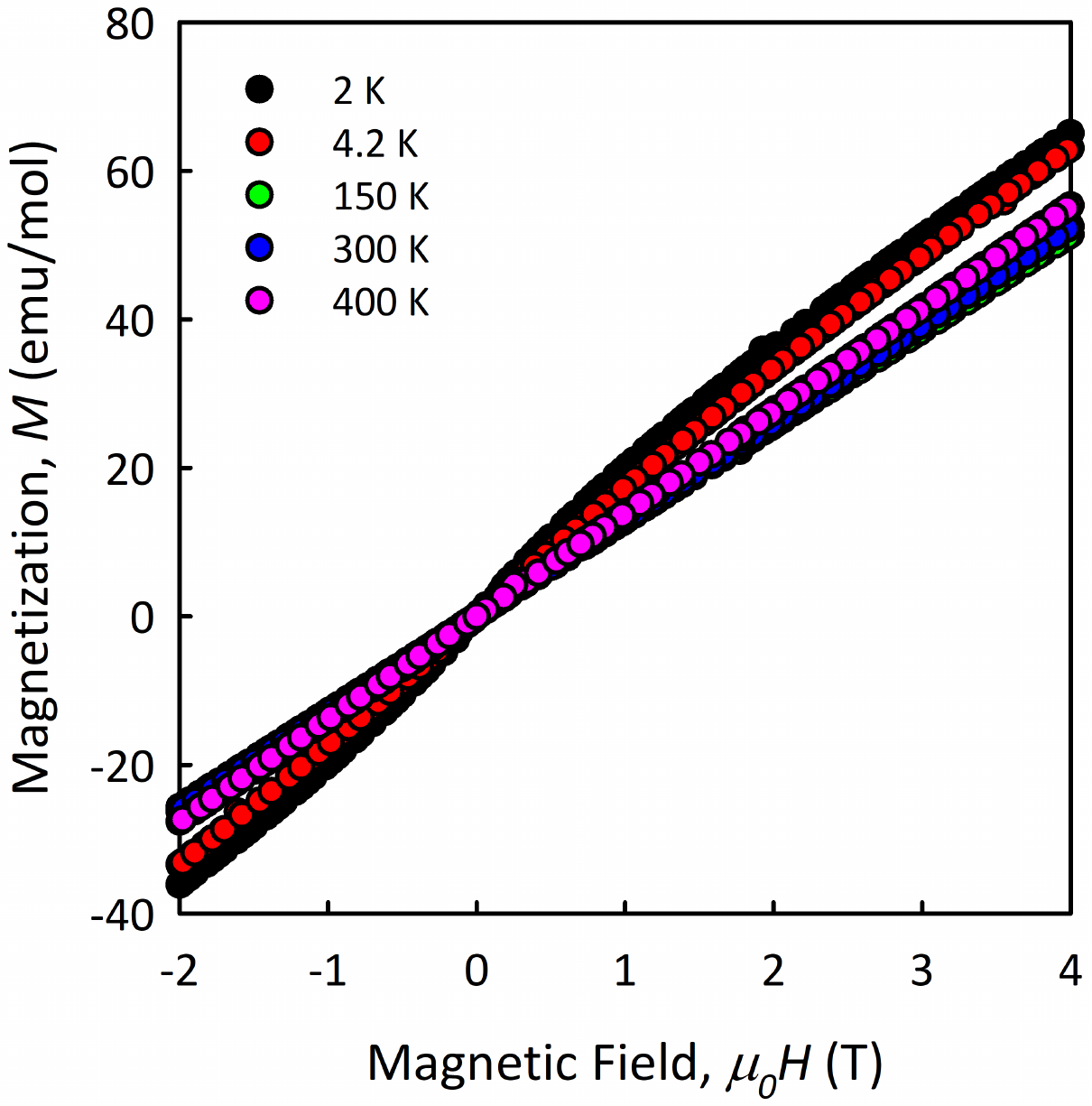}
\caption{Magnetization curves for BaCr$_2$P$_2$.}
\label{m_vs_h}
\end{figure}

{\it Magnetic Characterization.}
Magnetization was measured both as a function of temperature for a fixed field and 
as a function of field for various fixed temperatures spanning the range of 
the capabilities of our magnetometer (i.e. $1.9-400$ K). We present 
the zero-field cooled (ZFC) molar susceptibility in Fig. \ref{chi_vs_T}.
A large paramagnetic signature is present
at very low temperatures,
likely the result of a small quantity
of magnetic impurities being present. 
Fisher's analysis\cite{fisher_62} of $d(T \chi)/dT$ was used to determine 
the onset of the two transitions apparent in Fig. \ref{chi_vs_T}.
We find that the first transition (T1) is at $\sim 60$ K 
and that the second transition (T2) is at $342.4$ K. 
Magnetization versus magnetic field ($M(H)$)
measurements taken at different temperatures (Fig. \ref{m_vs_h})
are consistent with peak T1 being 
associated with an antiferromagnetic transition.
Further, the temperature dependence shown by the magnetization in Fig. \ref{chi_vs_T}
is similar to that shown by
related iron-pnictide compounds\cite{saparov_sefat_14,sefat_08,tegel_08} 
BaFe$_2$As$_2$, SrFe$_2$As$_2$, CaFe$_2$As$_2$, and SrFeAsF,
which also exhibit an antiferromagnetic groundstate.
The behavior whereby $\chi$ increases with increasing temperature 
was qualitatively explained in a simple model by Sales et al. \cite{Sales_10} 
to be ultimately due to the multiband nature of these materials,
where both electron and holes are present at the Fermi surface. 
Interestingly, this model predicts that most semimetals 
should see an increase in the magnetic susceptibility with increasing temperature.

The peak T2 at higher temperature is consistent with high-temperature XRD data that  show 
a slight inflection in the evolution of the $c/a$ ratio at this temperature.  
(See Fig. \ref{xrd_temp}.)
Future experiments involving single crystals or neutron diffraction measurements will serve 
to clarify the nature of the magnetism in this material. However, on the basis of our analysis on 
powder samples, combined with the literature on structurally similar compounds and 
our DFT calculations above, we posit that the compound BaCr$_2$P$_2$
is a G-type antiferromagnet,
with a N\'eel temperature of $\sim 60$ K. 

\begin{figure}
\includegraphics[scale=0.65]{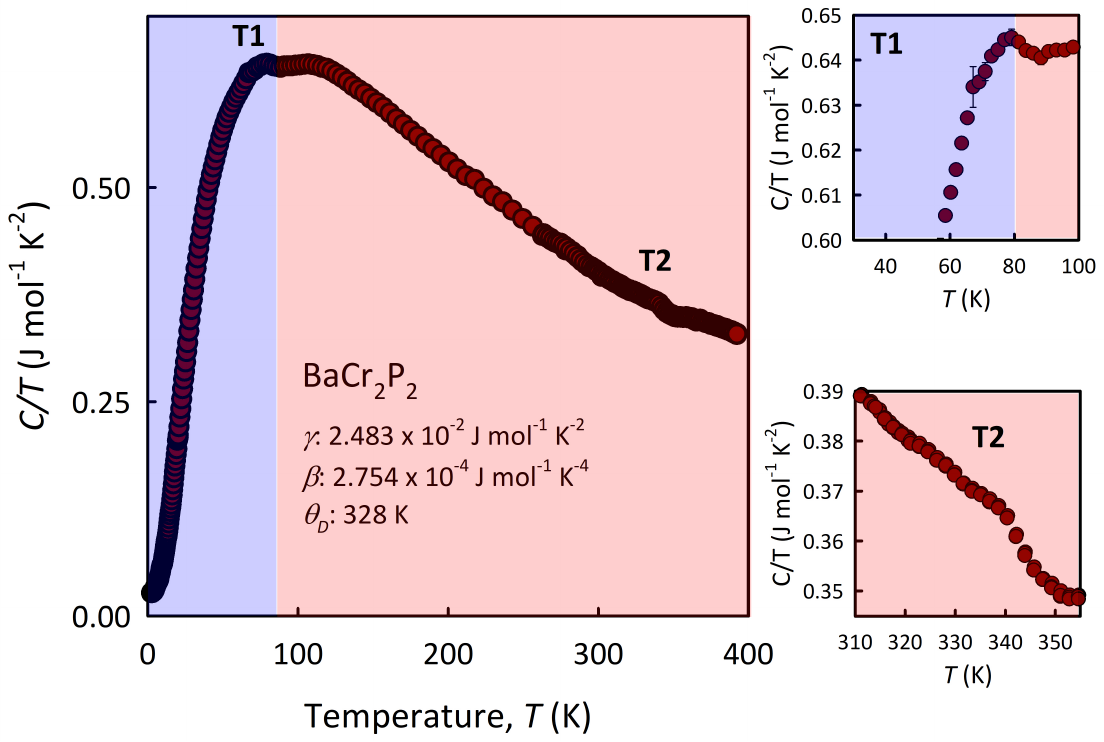}
\caption{Specific heat versus temperature for the compound BaCr$_2$P$_2$. 
The  transitions T1 and T2 are revealed in the two figures on the right-hand panel.}
\label{Cp_vs_T}
\end{figure}

{\it Thermodynamic Characterization.} 
Calorimetric measurements were taken via two different complementary techniques 
as described in the procedures section. 
The resulting data are presented in Fig. \ref{Cp_vs_T},
and they are consistent with the data from magnetization measurements 
-- both transitions T1 and T2 are again present, here at $\sim 75$ K and at $352$ K, 
in reasonable agreement with the same transitions noted in Fig. \ref{chi_vs_T}. 
Both the high-temperature and low-temperature data are noted in the smaller graphs to 
the side of the main plot. The data from $2.2$ to $20$ K was fitted to the equation
$C/T = \gamma + \beta T^2$  to elucidate the Sommerfeld coefficient 
($\gamma = 24.8$ mJ mol$^{-1}$ K$^{-2}$)  and the Debye temperature 
($\theta_D =  [1944/\beta({\rm mole-atom})]^{1/3} = 328$ K) of the compound BaCr$_2$P$_2$. 
For comparison, 
the Sommerfeld coefficient
$\gamma$ for the related G-type antiferromagnet BaCr$_2$As$_2$
has been reported with similar values\cite{filsinger_17,singh_09}:
$18.8$-$19.31$ mJ mol$^{-1}$ K$^{-2}$,
with a Debye temperature of $268$ K.

\section{Two-orbital hopping model}
In section \ref{DFT},
we reported that our DFT calculations
predict that the groundstate of BaCr$_2$P$_2$
is a  N\'eel antiferromagnet over the chromium atoms.
The same prediction was made by
Singh and coworkers in the case of BaCr$_2$As$_2$ \cite{singh_09}.
Both results raise the following question.  What is the physics that
underlies the G-type  antiferromagnetism in chromium pnictides predicted by DFT?
Below, we introduce a tight-binding model that provides some
insight into this question.

Our DFT calculation predicts that the electrons at the Fermi level
have roughly $2/3$ $d$-orbital character and $1/3$ $p$-orbital character.
Importantly, however,
the magnetic moment lies exclusively on the chromium site,
at which the electron is in the $3d$ orbital.
We believe, therefore, that it is sufficient
to include only the chromium $3d$ orbitals in order to describe magnetism
in chromium pnictides.
With this aim in mind,
we introduce the following model
for an electron that hops over a square lattice of chromium atoms
that includes only the principal $3d_{xz}$ and $3d_{yz}$ orbitals:
\begin{equation}
H_{\rm hop} =
-\sum_{\langle i,j \rangle} (t_1^{\alpha,\beta} c_{i, \alpha,s}^{\dagger} c_{j,\beta,s} + {\rm h.c.})
-\sum_{\langle\langle  i,j \rangle\rangle} (t_2^{\alpha,\beta} c_{i, \alpha,s}^{\dagger} c_{j,\beta,s} + {\rm h.c.}).
\label{hop}
\end{equation}
We work, specifically, in the isotropic basis of orbitals\cite{jpr_rm_18}
$d-=(d_{xz}-id_{yz})/{\sqrt 2}$ and $d+=(d_{xz}+id_{yz})/{\sqrt 2}$.
Above,
$c_{i, \alpha,s}$ and $c_{i, \alpha,s}^{\dagger}$
denote annihilation and creation operators for an
electron of spin $s$ in orbital $\alpha$ at site $i$.
Repeated indices are summed over.
Also above, $\langle i,j\rangle$ and $\langle\langle i,j\rangle\rangle$
represent nearest neighbor (1) and next-nearest neighbor (2) links on the
square lattice of chromium atoms.

The reflection symmetries shown by a single layer in a chromium pnictide
imply that the above intra-orbital and inter-orbital hopping matrix elements
show $s$-wave and $d$-wave symmetry, respectively\cite{raghu_08,Lee_Wen_08,jpr_mana_pds_14}.  
In particular,
nearest neighbor hopping matrix elements satisfy
\begin{eqnarray}
t_1^{\pm \pm} ({\bf {\hat x}}) &=& t_1^{\parallel} = t_1^{\pm \pm} ({\bf {\hat y}}) \nonumber\\
t_1^{\pm\mp} ({\bf {\hat x}}) &=& t_1^{\perp} = -t_1^{\pm \mp} ({\bf {\hat y}}),
\label{t1}
\end{eqnarray}
with real $t_1^{\parallel}$ and $t_1^{\perp}$,
while next-nearest neighbor hopping matrix elements satisfy
\begin{eqnarray}
t_2^{\pm \pm} ({\bf {\hat x}}+{\bf {\hat y}}) = \; t_2^{\parallel} &=& t_2^{\pm \pm} ({\bf {\hat y}}-{\bf {\hat x}}) \nonumber\\
t_2^{\pm \mp} ({\bf {\hat x}}+{\bf {\hat y}}) = \pm t_2^{\perp} &=& -t_2^{\pm \mp} ({\bf {\hat y}}-{\bf {\hat x}}),
\label{t2}
\end{eqnarray}
with real $t_2^{\parallel}$ and pure-imaginary $t_2^{\perp}$.

The above hopping Hamiltonian is easily diagonalized by
plane waves of $d_{x(\delta)z}$ and $i d_{y(\delta)z}$ orbitals that are rotated
with respect to the principal axes by a phase shift $\delta({\bm k})$:
\begin{eqnarray}
|{\bm k}, d_{x(\delta)z}\rangle\rangle &=&
{\cal N}^{-1/2} \sum_i e^{i{\bm k}\cdot{\bm r}_i}
[e^{i\delta({\bm k})} |i, d+\rangle + e^{-i\delta({\bm k})} |i, d-\rangle], \nonumber\\
i|{\bm k}, d_{y(\delta)z}\rangle\rangle &=&
{\cal N}^{-1/2} \sum_i e^{i{\bm k}\cdot{\bm r}_i}
[e^{i\delta({\bm k})} |i, d+\rangle - e^{-i\delta({\bm k})} |i, d-\rangle],
\label{plane_waves}
\end{eqnarray}
where ${\cal N} = 2 N_{\rm Cr}$ is the number of chromium site-orbitals.
Their energy eigenvalues are respectively given by
$\varepsilon_+({\bm k}) = \varepsilon_{\parallel}({\bm k}) + |\varepsilon_{\perp}({\bm k})|$ and
$\varepsilon_-({\bm k}) = \varepsilon_{\parallel}({\bm k}) - |\varepsilon_{\perp}({\bm k})|$,
where
\begin{eqnarray}
\varepsilon_{\parallel}({\bm k}) &=& -2 t_1^{\parallel} (\cos k_x a + \cos k_y a)
-2 t_2^{\parallel} (\cos k_+ a + \cos k_- a) \nonumber \\
\varepsilon_{\perp}({\bm k}) &=& -2 t_1^{\perp} (\cos k_x a - \cos k_y a)
-2 t_2^{\perp} (\cos k_+ a - \cos k_- a),
\label{mtrx_lmnts}
\end{eqnarray}
are diagonal and off-diagonal matrix elements,
with  $k_{\pm} = k_x \pm k_y$.
The phase shift $\delta({\bm k})$ is set by
$\varepsilon_{\perp}({\bm k}) = |\varepsilon_{\perp}({\bm k})| e^{i 2 \delta({\bm k})}$.
Specifically,
\begin{eqnarray}
\label{c_2dlt}
\cos\,2\delta({\bm k}) &=& {-t_1^{\perp}(\cos\, k_x a - \cos\, k_y a)\over
{[t_1^{\perp 2}(\cos\, k_x a - \cos\, k_y a)^2 +
|2 t_2^{\perp}|^2 (\sin\, k_x a)^2 (\sin\, k_y a)^2]^{1/2}}}, \nonumber \\
\label{s_2dlt}
\sin\,2\delta({\bm k}) &=& {2 (t_2^{\perp} / i)(\sin\, k_x a) (\sin\, k_y a)\over
{[t_1^{\perp 2}(\cos\, k_x a - \cos\, k_y a)^2 +
|2 t_2^{\perp}|^2 (\sin\, k_x a)^2 (\sin\, k_y a)^2]^{1/2}}}.
\label{phase_shift}
\end{eqnarray}
The phase shift is singular at ${\bm k} = 0$ and $(\pi/a,\pi/a)$,
where the matrix element $\varepsilon_{\perp}({\bm k})$ vanishes.

\begin{figure}
\includegraphics[scale=0.6]{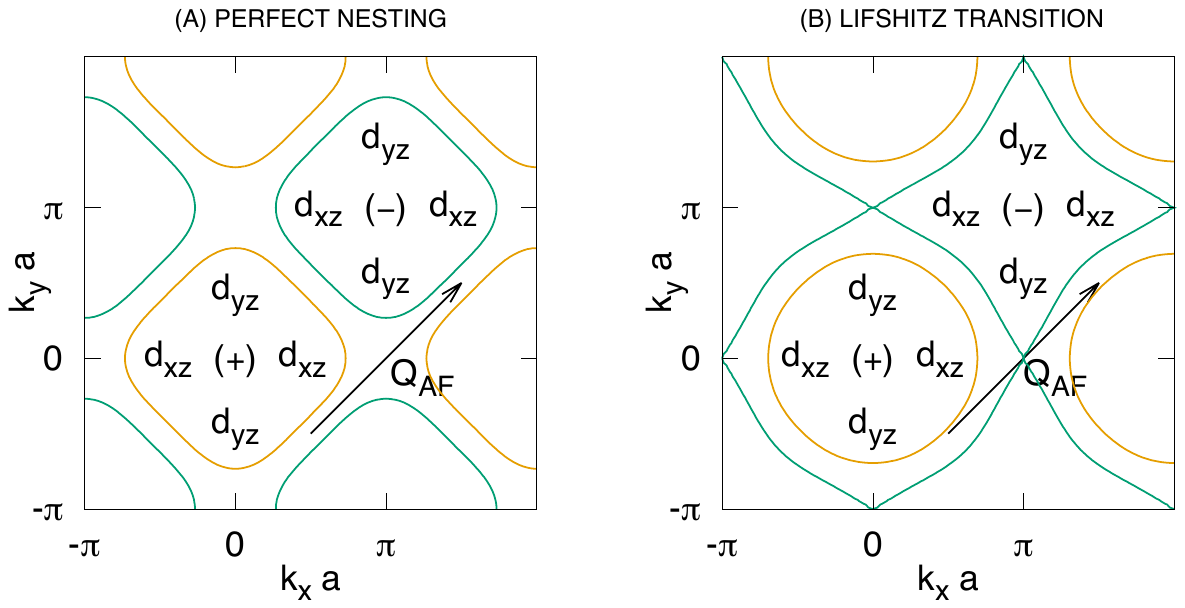}
\caption{(a) Perfect nesting at wavenumber ${\bm Q}_{\rm AF} = (\pi/a,\pi/a)$, with hopping parameters
$t_1^{\parallel} = -500$ meV, $t_1^{\perp} = -100$ meV,
$t_2^{\parallel} = 0$ and $t_2^{\perp} =  -100\, i$ meV; 
(b) residual nesting at Lifshitz transition, with 
$t_2^{\parallel} = -170$ meV instead.  The ``$+$'' and ``$-$'' symbols mark hole-type and electron-type
Fermi surfaces, respectively. Both (a) and (b) are at half filling.}
\label{FS0}
\end{figure}

If we first turn off next-nearest neighbor intra-orbital hopping,
$t_2^{\parallel} = 0$,
notice that the above energy bands then satisfy the perfect nesting condition
\begin{equation}
\varepsilon_{\pm}({\bm k}+{\bm Q}_{\rm AF}) = - \varepsilon_{\mp}({\bm k}),
\label{prfct_nstng}
\end{equation}
where ${\bm Q}_{\rm AF} = (\pi/a,\pi/a)$ is the N\'eel ordering vector on the square
lattice of chromium atoms. As a result, the Fermi level
of the bands lies at $\varepsilon_{\rm F} = 0$ at half filling.
Figure \ref{FS0}a shows such
perfectly nested hole-type ($+$) and electron-type ($-$) Fermi surfaces for hopping
parameters $t_1^{\parallel} = -500$ meV, $t_1^{\perp} = -100$ meV,
$t_2^{\parallel} = 0$ and $t_2^{\perp} =  -100\, i$ meV.
Figure \ref{FS0}b, on the other hand, shows residual nesting of the Fermi surfaces at half filling,
with hopping parameters 
$t_1^{\parallel} = -500$ meV, $t_1^{\perp} = -100$ meV,
$t_2^{\parallel} = -170$ meV and $t_2^{\perp} =  -100\, i$ meV.
Next-nearest neighbor intra-orbital hopping $t_2^{\parallel}$
has been tuned to a Lifshitz transition
at which open Fermi surfaces appear.  Increasing the strength
of $t_2^{\parallel}$ past this point results in two hole-type Fermi surfaces
at the center of the $1$-chromium Brillouin zone.
Such Fermi surface tubes notably resemble the outer tubular Fermi surfaces
obtained by DFT that are displayed by Fig. \ref{FS}.

The perfectly nested Fermi surfaces shown
by Fig. \ref{FS0}a can result in an instability
to long-range N\'eel  order.
Consider, in particular,
the spin magnetization
for magnetic order at wave number ${\bm Q}$ defined by
\begin{equation}
S_z ({\bm Q}) = \sum_i\sum_{\alpha} e^{i{\bm Q}\cdot{\bm r}_i} {1\over 2} (n_{i,\alpha,\uparrow} - n_{i,\alpha,\downarrow}).
\label{Sz}
\end{equation}
Here $n_{i,\alpha,s}$ measures the occupation of an electron 
of spin $s$ in orbital $\alpha$ at site $i$.
In momentum space,
it takes the form
\begin{eqnarray}
S_z ({\bm Q}) = {1\over 2}\sum_s\sum_{\bm k}\sum_{n,n^{\prime}}
({\rm sgn}\, s) {\cal M}_{n,{\bm k};n^{\prime},{\bm k}+{\bm Q}}\,
c_s^{\dagger}(n^{\prime},{\bm k}+{\bm Q}) c_s(n,{\bm k}),\nonumber\\
\label{sz}
\end{eqnarray}
where $c_s(n,{\bm k})$ and $c_s^{\dagger}(n,{\bm k})$
are annihilation and creation operators for an electron in
eigenstates (\ref{plane_waves}) of the
electron hopping Hamiltonian, $H_{\rm hop}$.
In particular,
$n=1$ and $2$ index the anti-bonding and bonding orbitals
$i d_{y(\delta)z}$ and $d_{x(\delta)z}$.
The above matrix element is computed in ref. \cite{jpr_rm_18}.
Importantly,
at the wave number that corresponds to N\'eel order,
${\bm Q}_{\rm AF}=(\pi/a,\pi/a)$,
 it is given by
\begin{equation}
{\cal M}_{n,{\bm k};n^{\prime},{\bm k}+{\bm Q}_{\rm AF}} =
\begin{cases}
\pm \sin 2\delta({\bm k})  & \text{for}\quad n^{\prime} = n , \\
\pm i \cos 2\delta({\bm k})  & \text{for}\quad n^{\prime} \neq n .
\end{cases}
\label{M}
\end{equation}
The contribution to the static spin susceptibility from inter-band scattering
that corresponds to 
N\'eel order
is then given by the Lindhard function
\begin{equation}
\chi_{\rm inter}({\bm Q}_{\rm AF}) = -{1\over{a^2 N_{\rm Cr}}} \sum_{\bm k}
{n_F[\varepsilon_-({\bm k}+{\bm Q}_{\rm AF})] - n_F[\varepsilon_+({\bm k})]\over
{\varepsilon_-({\bm k}+{\bm Q}_{\rm AF})-\varepsilon_+({\bm k})}}
|\cos 2\delta({\bm k})|^2 ,
\label{Lndhrd}
\end{equation}
where $n_F$ is the Fermi-Dirac distribution.  Applying the perfect-nesting
condition (\ref{prfct_nstng}) yields the more compact expression
\begin{equation}
\chi_{\rm inter}({\bm Q}_{\rm AF}) = {1\over{a^2 N_{\rm Cr}}} \sum_{\bm k}
{{1\over 2} - n_F[\varepsilon_+({\bm k})]
\over
{\varepsilon_+({\bm k})}}
|\cos 2\delta({\bm k})|^2 .
\label{prfct_Lndhrd}
\end{equation}
We conclude that the static susceptibility for N\'eel order diverges
logarithmically as
$\chi_{\rm inter}({\bm Q}_{\rm AF}) = {\rm lim}_{\epsilon\rightarrow 0}\, c^2D_+(0)\, {\rm ln}(W_{\rm bottom}/\epsilon)$,
with corresponding density of states weighted by the
magnitude square of the matrix element (\ref{M}):
\begin{equation}
c^2D_+(\varepsilon) = (2\pi)^{-2} \int_{1\rm Cr\: BZ}d^2 k\, [\cos 2\delta({\bm k})]^2\delta[\varepsilon-\varepsilon_+({\bm k})] .
\label{cD}
\end{equation}
Above, $W_{\rm bottom} = - \varepsilon_+(\pi/a,\pi/a)$.
The logarithmic divergence of $\chi_{\rm inter}({\bm Q}_{\rm AF})$ 
at perfect nesting guarantees an instability towards long-range N\'eel order.

\begin{figure}
\includegraphics[scale=0.5]{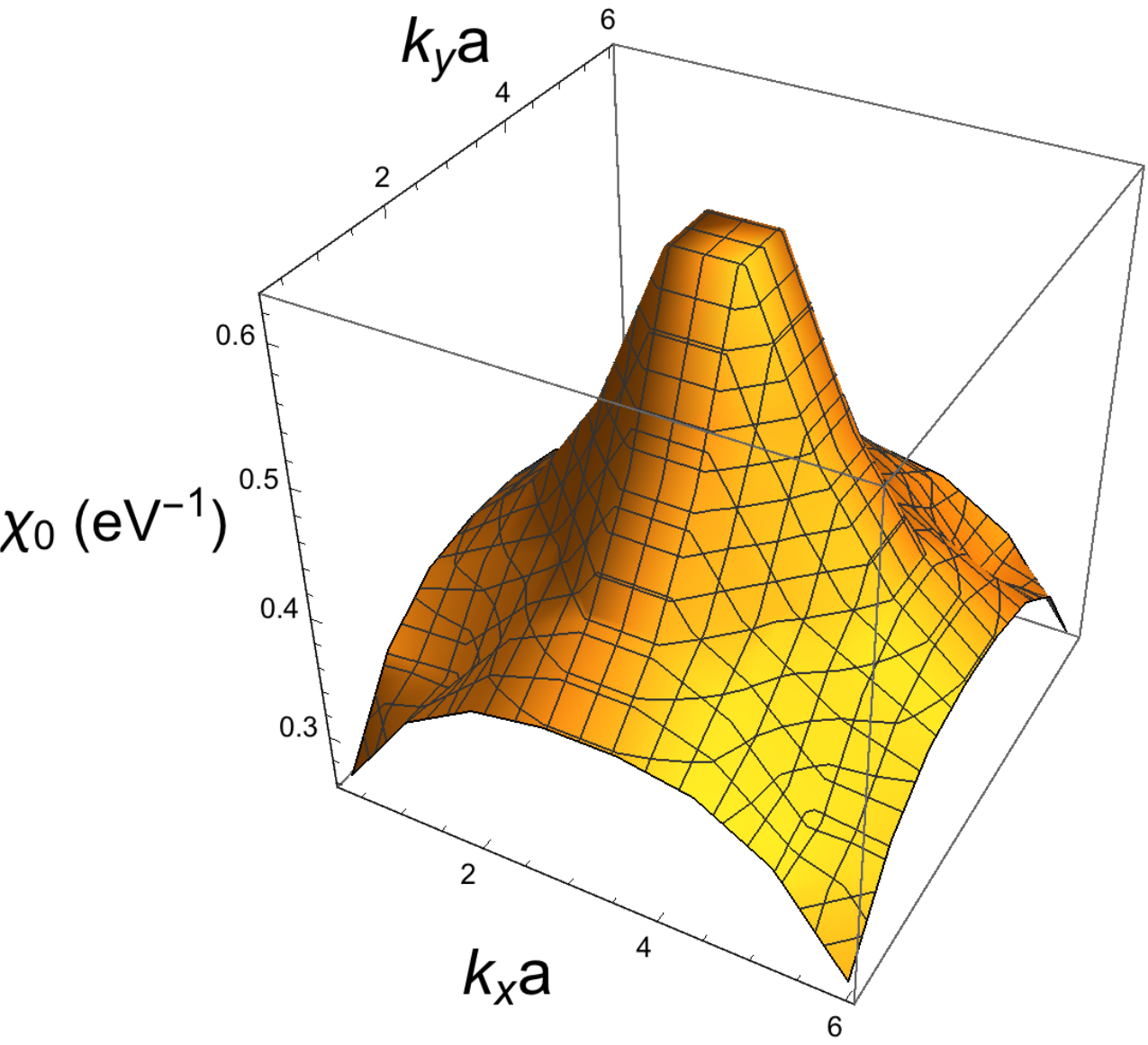}
\caption{Static spin susceptibility  at half filling and at perfect nesting
over a $20\times 20$ lattice of chromium atoms.
Electron hopping parameters coincide with those in Fig. \ref{FS0}a.}
\label{chi0_prfct}
\end{figure}

\begin{figure}
\includegraphics[scale=0.5]{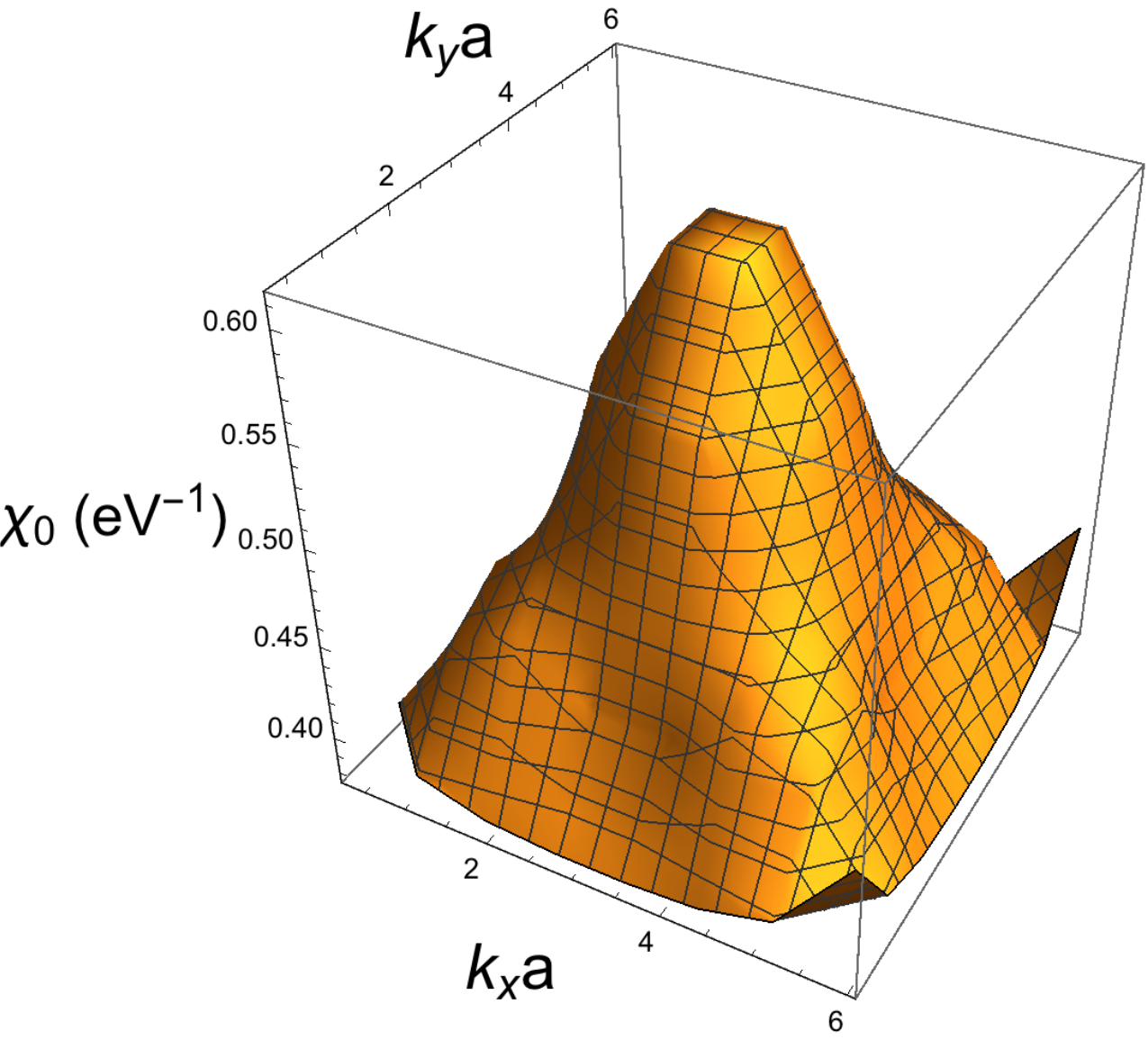}
\caption{Static spin susceptibility  at half filling and at the Lifshitz transition
over a $20\times 20$ lattice of chromium atoms.
Electron hopping parameters coincide with those in Fig. \ref{FS0}b.}
\label{chi0_lfshtz}
\end{figure}

Yet can Fermi surfaces that show residual nesting,
such as the one displayed by Fig. \ref{FS0}b,
also support an instability towards N\'eel order?
To answer this question, we shall compute the static magnetic susceptibility
at all wavenumbers.  In general, the intra-band and inter-band susceptibilities
take the form
\begin{equation}
\chi_{n,n^{\prime}}({\bm Q}) = -{1\over{a^2 N_{\rm Cr}}} \sum_{\bm k}
{n_F[\varepsilon_{n^{\prime}}({\bm k}+{\bm Q})] - n_F[\varepsilon_n({\bm k})]\over
{\varepsilon_{n^{\prime}}({\bm k}+{\bm Q})-\varepsilon_{n}({\bm k})}}
|{\cal M}_{n,{\bm k};n^{\prime},{\bm k}+{\bm Q}}|^2.
\label{LNDHRD}
\end{equation}
The matrix element above has been computed in ref. \cite{jpr_rm_18},
and it's magnitude square is given by
\begin{equation}
|{\cal M}_{n,{\bm k};n^{\prime},{\bm k^{\prime}}}|^2 =
\begin{cases}
\cos^2[\delta({\bm k})-\delta({\bm k^{\prime}})] & \text{for}\quad n^{\prime} = n ,\\
\sin^2[\delta({\bm k})-\delta({\bm k^{\prime}})] & \text{for}\quad n^{\prime} \neq n .
\end{cases}
\label{A_M2}
\end{equation}
Next, the latter can be written in terms of the expressions (\ref{phase_shift})
for ${\rm cos}(2\delta)$ and ${\rm sin}(2\delta)$
by the application of half-angle formulas.
The net static spin susceptibility is then
$\chi_0({\bm Q}) = {1\over 2}\sum_{n=1,2}\sum_{n^{\prime}=1,2} \chi_{n,n^{\prime}}({\bm Q})$.
Figures \ref{chi0_prfct} and \ref{chi0_lfshtz} 
show the static spin susceptibility 
at perfect nesting and at the Lifshitz transition.  
Electron hopping parameters are identical to those in Figs. \ref{FS0}a and \ref{FS0}b.
Both cases notably show a well-defined peak at ${\bm Q}_{\rm AF}$.  
The logarithmic singularity predicted in the case of perfect
nesting is clearly missing in Fig. \ref{chi0_prfct}
because the thermodynamic limit has not been achieved.

In summary, the present two-orbital hopping model (\ref{hop})
can describe Fermi surface tubes
at the center of the $1$-chromium Brillouin zone that resemble qualitatively the
outer tubular Fermi surfaces obtained by DFT. (Cf. Figs. \ref{FS} and \ref{FS0}b.)  
Such Fermi surfaces show residual nesting by the N\'eel wavenumber when they are nearby
a Lifshitz transition to open Fermi surfaces. This suggests that similar physics
underlies our DFT prediction of N\'eel antiferromagnetism in chromium pnictides.

\section{Summary and Conclusions}
After performing DFT calculations on the new chromium-pnictide compound BaCr$_2$P$_2$,
we predict that it has the ThCr$_2$Si$_2$ crystal structure\cite{xtal_strctr}, 
in common with its sister chromium-pnictide 
compound\cite{singh_09,filsinger_17} BaCr$_2$As$_2$.
We also successfully synthesized a powder sample of the new material.  
XRD analysis of the sample yields the predicted crystal structure to within $1$ percent accuracy.
Our DFT calculations also predict N\'eel antiferromagnetism in BaCr$_2$P$_2$,
with magnetic moments that lie primarily on the chromium atoms.
Magnetic susceptibility and specific-heat measurements versus temperature show a kink
near $60$ K that we tentatively attribute to the transition temperature for
the predicted antiferromagnetic state.
Last, by comparison with a simple tight-binding model that contains only the principal
$3d_{xz}$ and $3d_{yz}$ orbitals, we suggest that the N\'eel antiferromagnetic order predicted
by DFT is a result of residual nesting of the outer tubular Fermi surfaces 
that is obscured by a Lifshitz transition.
(Cf. Figs. \ref{FS} and \ref{FS0}b.)

Unlike iron pnictides,
no traces of superconductivity in un-doped or doped chromium pnictides 
have yet been reported in the literature.
Like iron pnictides, however, 
DFT predicts antiferromagnetic order in chromium-pnictide parent compounds.
As mentioned above, it's quite possible that the previous is due to 
weak nesting of the tubular Fermi surfaces by the N\'eel wavenumber.  
By analogy with the nesting of the
hole-type and electron-type Fermi surfaces that exists in parent compounds to
iron-pnictide superconductors,  a related type of superconductivity may lurk
in chromium-pnictide materials that are suitably doped.
Indeed, the peak in the static spin susceptibility at the N\'eel wavevector
displayed by Fig. \ref{chi0_lfshtz}
indicates that antiferromagnetic spin fluctuations are present.
By analogy with theoretical predictions
for the nature of superconductivity in iron-pnictides\cite{mazin_08,kuroki_08},
such spin fluctuations can result in $S$-wave Cooper pairing
that alternates in sign between the two tubular Fermi surfaces.
(See Figs. \ref{FS} and \ref{FS0}b.)
These theoretical considerations suggest looking for superconductivity in doped BaCr$_2$P$_2$
as well.

\begin{acknowledgments}
The authors thank Dr. Yoshihisa Ichihara for valuable discussions.
This work was supported in part by the US Air Force
Office of Scientific Research (AFOSR) under grant no. FA9550-17-1-0312
and by the National Science Foundation under PREM grant no. DMR-1523588
and CREST grant no. HRD-1547723.
It was also supported in part by the Aerospace Systems Directorate at the Air Force Research Laboratory,
by AFOSR grant LRIR 18RQCOR100, and by a grant from the National Research Council.
\end{acknowledgments}




\end{document}